# VEHICLE COLLISION DETECTION & PREVENTION USING VANET BASED IOT WITH V2V

Shafin Talukder, SK. Tasnim Bari Ira, Aseya Khanom,
Prantika Biswas Sneha and Wardah Saleh

Department of Computer Engineering,
American International University- Bangladesh (AIUB), Dhaka, Bangladesh

## ABSTRACT

*EMERGENCY alert in case of any accident is vitally necessitated to rescue the victims. And so, this paper is made to present the results of a major analysis relating to emergency alert conditions at the time of collision (automobile). In this study, the authors have investigated modern Internet of Things (IoT) and VANET (Vehicular Ad hoc Networks) technologies and developed a collection of modern and specialized techniques as well as their characteristics. It has sensors that detect unbalanced circumstances and provide with a warning to the microcontroller if a collision occurs. Additionally, the technique can be implemented in such a way that vehicles are alerted of possible closing barriers. Vehicle-to-Vehicle communication (V2V) has a huge impact since it allows vehicles to communicate with each other while in proximity and the buzzer together with the LEDs serves as a safety feature. The primary goal of the system is to carry out the microcontroller functions in every environment and moreover, the concept refers to detect and prevent the collision specially in a foggy weather as well as at night and in other odd circumstances. The Internet of Things (IoT) and the Vehicular Ad-Hoc Network (VANET) have now been merged as the fundamental and central components of Intelligent Transportation System (ITS). Furthermore, while the procedure of obtaining the insurance may be longer for certain people. On the other hand, others may avoid the law after being involved in severe collisions which makes it difficult for the authorities to discriminate between criminal and non-criminal evidence.*

## KEYWORDS

## 1. INTRODUCTION

Collisions are one of the negative effects of any transportation system [1] and road accidents adversely impact developing countries on a regular schedule. The main reasons are inadequate infrastructure, traffic control, and accident management. South Asia, particularly India and Bangladesh are identified as the developing countries with the highest frequency of accidents [2]. However, technological superiority is in sight that exists in a universe whereby advanced technologies are also being developed and these approaches can be used in our society to fix shortcomings. At present the Internet of Things (IoT) is a figurative concept which depicts global internet connectivity [3]. The core idea behind the IoT concept is to spend billions if not trillions of smart devices which can assess the obtained data and detect any kind of collisions as well as general climate of the day of occurrence [4]. By the end of 2021 it is expected that there will be 28 billion connected devices [5] and IoT systems are a network that connects devices to collect and share data, and they are utilizedin a variety of applications [6]. An ad hoc network for automobiles is a network of moving vehicles where each vehicle acts as a node in the creation of a mobile network.

The number of causalities increases almost every year, and the ratio will never decrease if the trend continues to move in the same direction [7]. The table below represents a picture in terms of how dangerous the existing roads of Bangladesh are (Source: Bangladesh Police).

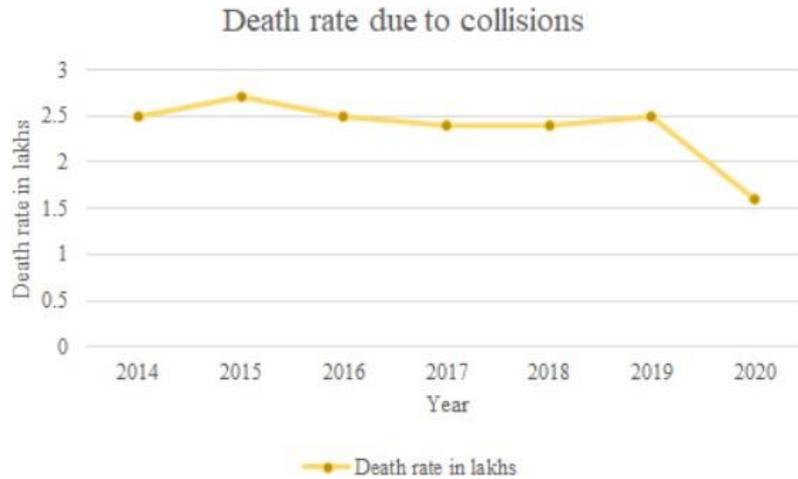

Figure 1. Total Estimated Number of Deaths Due to Accidents from 2014 to 2020[9].

## 2. LITERATURE REVIEW

In paper [8], the authors propose a GSM/GPRS/GPS-based car accident detection and rescue information system. It was developing a web service that alerts the owner of the vehicle and the nearest police station as well as the hospital about the incident together with its location [ 9 ]. The goal of this work is to create an IoT-based vehicle accident detection and rescue information system that can detect vehicle accidents and relay location information to the vehicle owner and nearby hospital as well as police station via a web service. The GSM/GPRS shield is used to establish a connection between the web server and the hardware device while the GPS shield is used to track the location. Vibration sensors, a keypad, and a buzzer detect the accident [10].

The proposed IoT-based car accident detection and notification algorithm for general road accidents aim to create a system that allows hospitals and other emergency platforms to receive real- time updates on nearby incidents. It is an IoT-based framework that is created to keep people updated on incidents so that emergency medical help may be provided with as earliest as possible. This can be achieved by utilizing smart sensors with a microprocessor within the vehicle that can be activated in the event of an accident [13].

The main goal of this paper is to detect over-speeding vehicles which is based on a speed limit and then alert the appropriate authority and concerned individuals by using an IoT-based framework [14]. A GPS module, radar, Google maps, and an IoT module are all vital part of the system; GPS and IoT technology are used to automatically control the safe zones. The battery performance of this activity tracking gadget is between 5 to 10 hours and the sensor's purpose is to effectively reduce accident death rates [15]. It is driven by 12V lithium batteries and has a GPS sensing network and IoT application.

Monitoring of traffic at night a robust framework for multi-vehicle detection, classification, and tracking shows how to convert headlights into images of a vehicle to evaluate the features of another vehicle viz [17]. with the screen or vehicle's shape and size. This is a two-stage system to detect with a tracking module that can handle partial and complete objects—the occlusion reasoning technique which takes stakes advantage of the headlamp positions and the basic traffic

34

scene layout. For scalability and to avoid application-specific cameras the system's characteristics are generic, so there is no need to change camera settings like low exposure [19].

## 3. PROPOSED METHOD

Collision can take place for several reasons. But according to the statistics of underdeveloped countries accidents occur mostly due to the foggy weather and lack of enough light at night. Consequently, it is proposed to develop a system that alerts people when an incident is potentially harmful if it takes place; therefore, the system ensures the best level of driving safety. The system quickly notifies the accidental area and in order to achieve this goal, a VANET based IoT vehicle is developed to detect accident and rescue the victims quickly. It also uses GPS and GSM to adjacent emergency hospitals, police stations, and an alert message is sent to the registered family members. On the other hand, VTS (Vehicle Tracking System) functions through the GPS to identify the nearest rescue platforms as well. Unlike most other systems this proposed work also discusses the concept of reporting collisions to the head offices of insurance company inorder to acquire relevant report of all documents linked to the specific vehicle.

## 4. VANET (VEHICULAR AD HOC NETWORKS)

A wireless ad hoc network (WANET) is a sort of on-demand, spontaneous device-to-device network. "Ad hoc" implies makeshift or impromptu. Without needing to connect to a Wi-Fi access point or router, we can set up a wireless connection straight to another computer or device in ad hoc mode. There are different types of ad hoc networks, but we are using VANET. The Vehicular Ad-hoc Network is essentially an infrastructure-free network. It enhances security approaches and convenience while driving, as well as allowing vehicles to communicate security information. The information shared in this system is time critical, necessitating the establishment of reliable and fast network connections. The major concern of VANET is impromptu networking. Facilities like RSUs (roadside units) and cell phone networks are less of an issue.

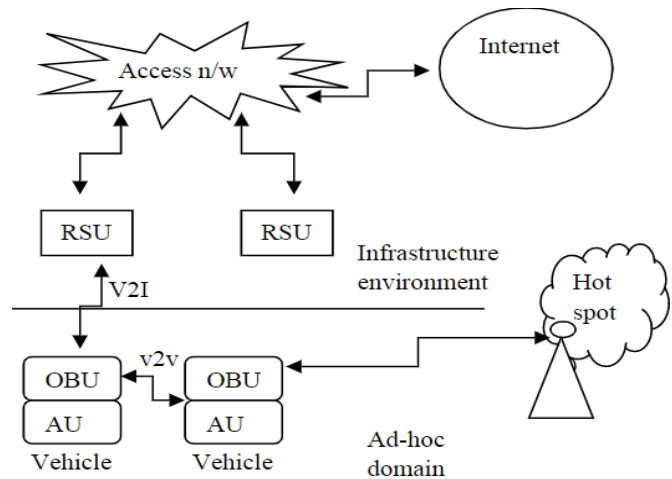

Figure 2. Architecture of VANET

## 5. LIST OF COMPONENT AND TECHNOLOGY

Based on a thorough evaluation of the papers and a step-by-step review of the results, the components and technologies which have chosen Ultrasonic Sensor, Vibration Sensor/

Accelerometer, LED, Buzzer, GPS, GSM, Push Button, Wi-Fi Module, Cloud Server and Arduino.

## 6. GRAPHICAL ARCHITECTURE OF REQUIREMENT TOOL

To address the scarcity of publicly available standard datasets for collision detection and prevention, an IoT-based accident detection and prevention system are suggested. The system architecture is depicted below. Most of the cars have now feature of built-in GPS devices to assist drivers.

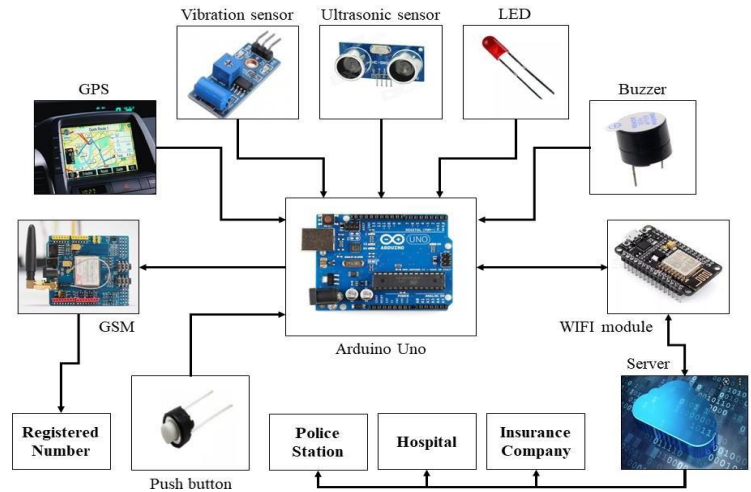

Figure 3. Graphical architecture

## 7. DESCRIPTION OF THE ARCHITECTURE'S CONNECTION

An IoT based accident detection and prevention system is proposed on emergency alert circumstances at the time of a collision. And therefore, emergency service is provided if any collision is detected. Figure 3 presents the system architecture; most automobiles now have the built-in GPS systems to assist passengers while driving. Under these circumstances, the vibration sensor and the ultrasonic sensor are both activate at the same moment. Basically, ultrasonic sensor is used to estimate the distance between the vehicles with the help of ultrasonic sound waves. The ultrasonic sensor in the proposed system is linked to an Arduino Uno microcontroller. This sensor is also used to measure distance between the vehicles and the nearby obstacles; the system then notifies the driver via red, green, yellow, and blue LEDs and the red LED alert accompanied by a buzzer sound. It should be mentioned that the LEDs are connected to the microcontroller and through V2V communication, an alert message is delivered to all vehicles within a certain range. It is to be noted that V2V communication only works for those vehicles which is equipped with the identical system. The vibration sensor is used to detect the unbalanced situation and it is connected to the Arduino Uno and when the vibration value exceeds the threshold limit, the sensor detects a mishap and alerts the Arduino-Uno. A single push button switch is connected to the microcontroller; the system considers the driver is in safe condition and does not advance to the following phase if the button is pressed within 30 seconds. On the other hand, if the button is not pressed in the specific time, then the system is advanced to the next phase because it considers the driver is in serious condition. The data is subsequently sent via the processor's Wi-Fi module to the cloud server and the microcontroller is linked to the Wi-Fi module in the suggested system. Simultaneously, the system sends SMS to the registered number by using the GSM module and in the system GSM Module is connected to the Arduino-uno. The cloud server scans the database for phone numbers of the nearest hospitals, police stations, and insurance companies. As a result, hospital dispatches an ambulance to the accidental place along with a police officer for investigation. The reports of the collision will also be given to the insurance companies so that their representative can visit the spot and process the insurance claim procedure accordingly.

## 8. FLOW CHART

In this segment a flowchart encompassing all the specifications and instructions are being demonstrated. Each functions respond differently when it comes to the system's activities under each circumstance and those functions are defined accordingly.

The microprocessor in the system is Arduino, begins its' functionality as soon as the ignition of the vehicle gets on. Two sensors are used in the whole system: an ultrasonic sensor and a vibration sensor with their own set of activities. Both the sensors begin working after the ignition gets started and carry out their functions accordingly.

Firstly, the system measures the distance between the vehicles and the nearby obstacles around it every second after the Arduino's reading of data from the ultrasonic sensor. A green LED blinks continuously if there is no vehicle within "50 meters" of the system. Nevertheless, the system begins flashing with a yellow LED if any object or vehicle enters within that range of "50 meters". The yellow LED continues to flash until the distance is between "30 to 50 meters". After that if any object or vehicle the appears within the 30 meters, a red LED begins blinking. The system also works with a warning message via V2V communication. The most critical part is a buzzer which starts making sounds to confirm the probability of a collision if any vehicle or object appears within 10 meters.

Following that if a collision occurs, the sensor confirms the collision, and the GPS detects the vehicle's location immediately. There is also a push-button function in the system that the driver can use, and the system does not send any alarm messages if the driver presses that button within the first 30 seconds. However, the system will proceed to the next step if the button is not pressed within the first 30 seconds which is the GSM and Wi-Fi modules get activated at the same time. The Wi-Fi module searches and delivers the data to the server. By following the gathered data, the Wi-Fi module looks for the phone numbers of nearby hospitals, police stations, and insurance companies. GSM on the other hand, delivers the alert message with the necessary information to the registered family phone numbers.

On the other hand, the vibration sensor checks for any disproportionate condition and if it detects one, then the system blinks a blue LED to inform the driver about the situation. The system proceeds from the scratch if the driver presses the push button within the first 30 seconds but if the button is not pressed then GPS monitors the location and the GSM as well as Wi-Fi modules execute their functions as before.

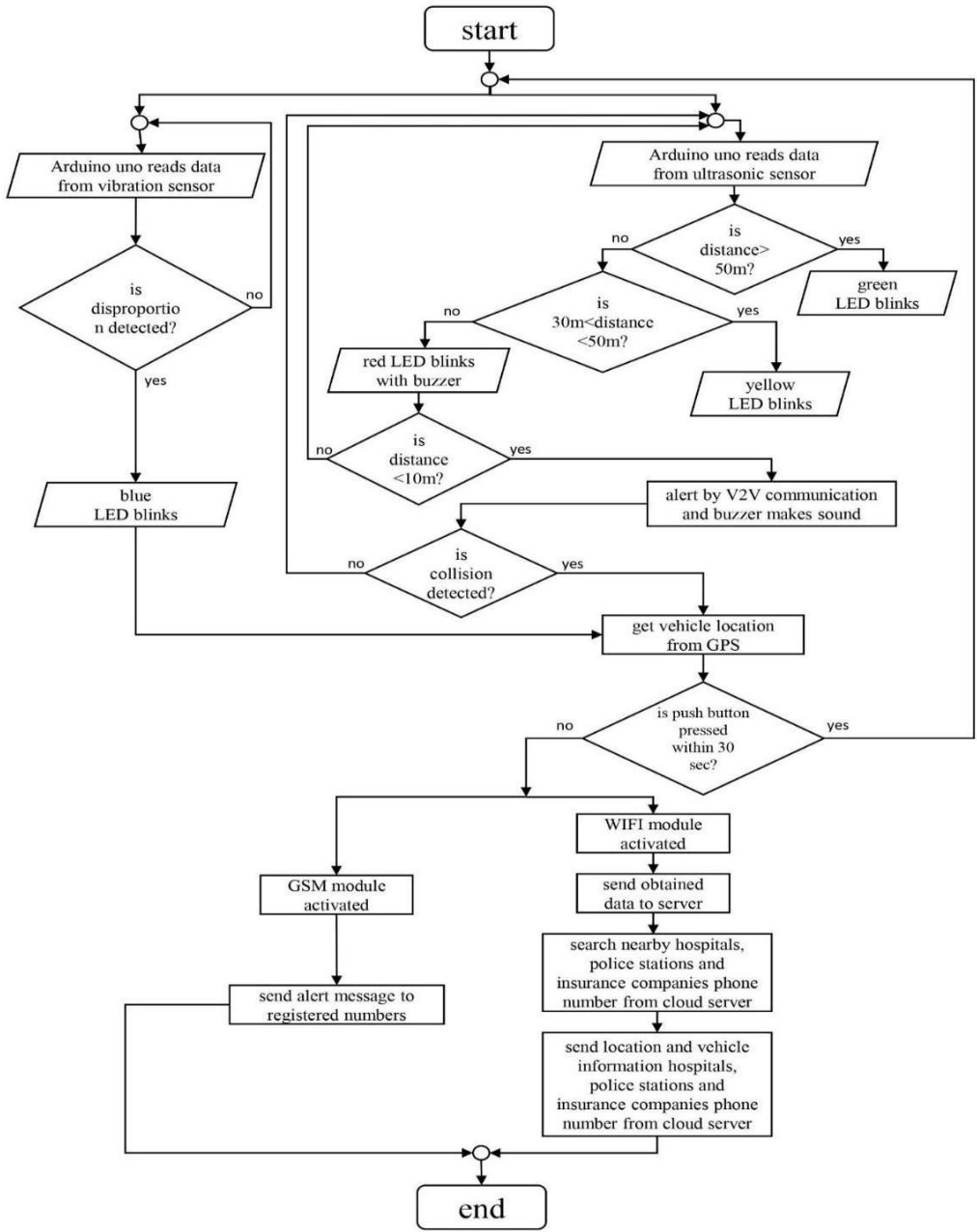

Figure 4. Flow chart

# 9. CASE DIAGRAM

The following picture shows the case diagram of the system; the case diagram helps to understand the connectivity between the actors and the actions. The following one represents the relations between the user and the system of the proposed system.

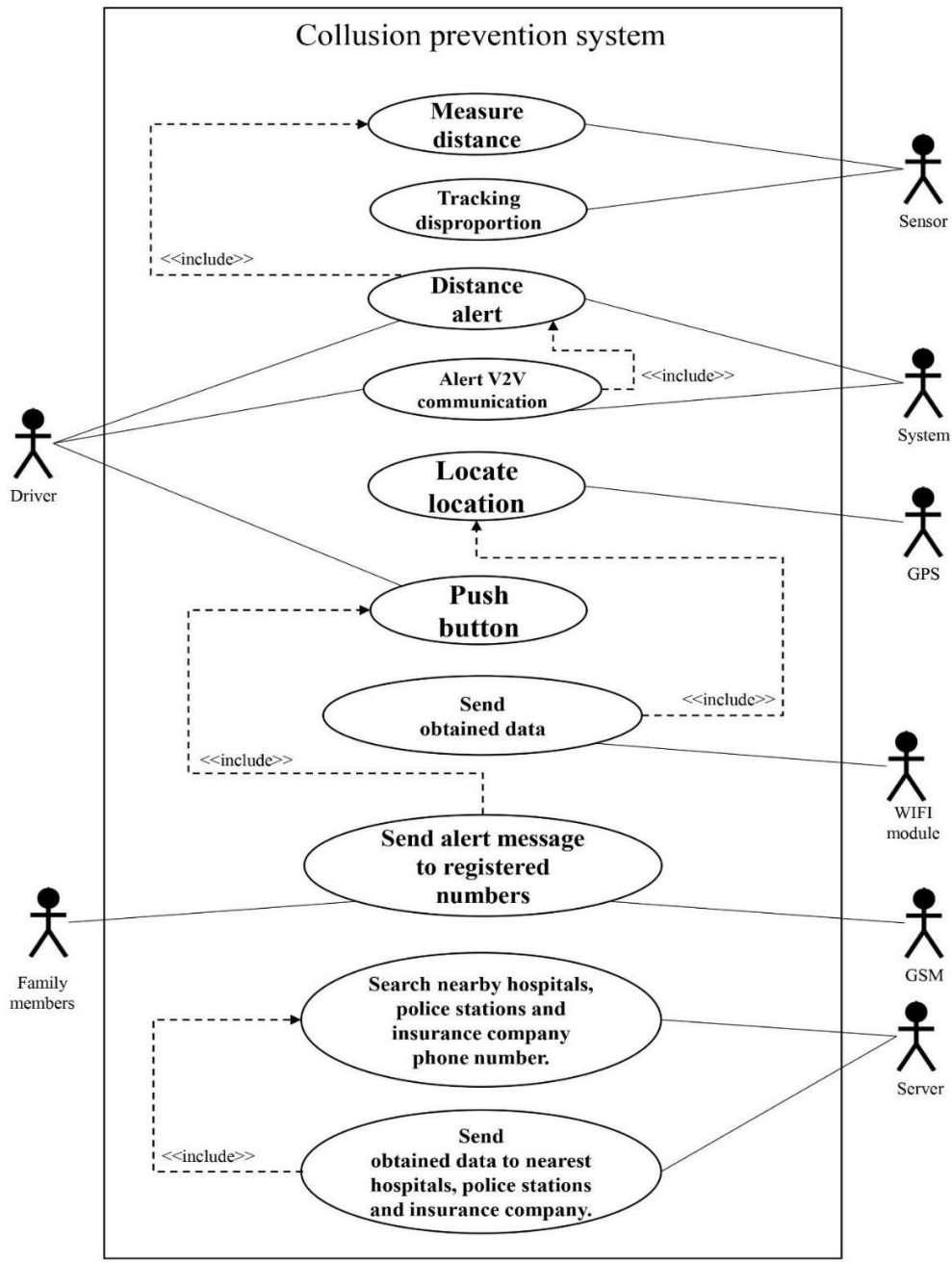

Figure 5. Case diagram

## 10. RESULT

The table shows the different types of solutions from previous related works. The system is designed which is based on the vehicle collision problem that works in all different climates including day and night. The table has been prepared which is based on the previous papers; the showcasing o f 8 most relevant papers with their features highlight why this system is better than the existing ones. It should also be noted that the proposed system not only ensures the driving safety but also helps to safeguard the lives by shortest possible time.

Table 1. Comparison Table

| Title reff | Location detection | Alert messages | Alert based on distance | Family member | Police station | Hospital | Insurance Company |
|---|---|---|---|---|---|---|---|
| [12] | ✓ | | | ✓ | ✓ | ✓ | ✓ |
| [18] | ✓ | | | | ✓ | ✓ | |
| [16] | | ✓ | ✓ | | | | |
| [08] | ✓ | | ✓ | ✓ | | | |
| [11] | ✓ | | | ✓ | | ✓ | |
| [21] | ✓ | ✓ | | ✓ | | ✓ | |
| [02] | ✓ | ✓ | | | | ✓ | |
| [20] | ✓ | ✓ | | | | ✓ | |
| Our System | ✓ | ✓ | ✓ | ✓ | ✓ | ✓ | ✓ |

## 11. ESTIMATED COST

This segment presents an estimated cost of the proposed system. As per following table the approximate cost of the entire proposal is calculated around BDT 2000 which is quite affordable.

Table 2. Table of the estimated cost

| Components | Quantity | EstimatePrice (BDT) | Estimate Price (Dollar $) |
|---|---|---|---|
| Ultrasonic sensor | 2 | 200 | 2.33 |
| Accelerometer | 1 | 279 | 3.25 |
| LED | 4 | 16 | 0.19 |
| Buzzer | 1 | 15 | 0.18 |
| Push Buttonswitch | 2 | 10 | 0.12 |
| Arduino Uno | 1 | 609 | 7.16 |
| Wi-Fi module | 1 | 410 | 4.82 |
| GSM Module | 1 | 403 | 4.74 |
| | **Total price** | **1942** | **22.64** |

## 12. DISCUSSION

The proposal is developed with the influence of rapid and advanced technology of VANET and IoT in terms of vehicle collisions. Unlike other proposals, this system can work in different situations and additionally the installation of the system is very convenient and easy to understand and use. Consequently, it encourages the vehicle owners to experience the using of this system. The use of two different sensors ensures the result of the highest level of driving safety and keep the passengers protected as well. And the technology of V2V is useful which can introduce an upper level of technology in under-developed countries like angladesh. The latest part of VTS also encourages a certain number of platforms to develop different fields of security; it also takes the technology to an advanced platform and this technology introduces an advanced use of those GPS which already exist in most of the cars running in under-developed countries like Bangladesh.

# 13. CONCLUSIONS

VANET and IoT are rapidly evolving technologies which are effectively implemented in vehicle-related systems. This proposed system is explored with the help of advanced technology. Undeniably, the provided solution has many advantages compared to previous ones. The most important part is that the proposed solution works for two separate circumstances and the technology assures the driver's safety by utilizing several types of LEDs for different length of distance. Additionally, V2V communication adds the advantage to warn the nearby vehicles in terms of the distance which is less than 10 meters in between the two vehicles. But it is also to be noted that the V2V communication only works if the other vehicle is also equipped with the same system. The system works with emergency rescue functionalities immediately if a collision occurs. Besides, it keeps monitoring the present state of the vehicle with the help of GPS. Compared to other systems this proposed approach is not costly at all. Even though the suggested system is designed simple, easy to use and flexible but still then there are few known problems which need to be resolved.

## AUTHORS

Name: **Shafin Talukder**
Mail: shafintalukder.nill@gmail.com Institute: American International niversity-Bangladesh (AIUB)

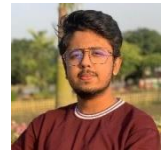

Name: **SK. Tasnim Bari Ira**
Mail: tasnimbariira@gmail.com Institute: American International University-Bangladesh (AIUB)

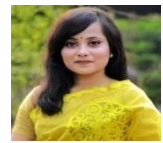

Name: **Aseya Khanom**
Mail: aseyakhanom99@gmail.com Institute: American International University-Bangladesh (AIUB)

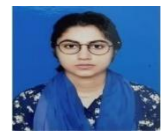

Name: **Prantika Biswas Sneha**
Mail: prantika.sneha@gmail.com Institute: American International University-Bangladesh (AIUB)

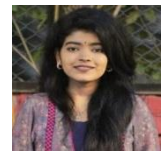

Name: **Wardah Saleh**
Mail: wardahsaleh15@gmail.com Institute: American International University-Bangladesh (AIUB)

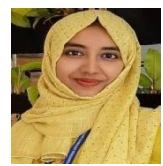